\documentclass[10pt,letterpaper]{article}
\usepackage[top=0.85in,left=1.in,footskip=0.75in,marginparwidth=2in]{geometry}

\usepackage[utf8]{inputenc}

\usepackage{cite}

\usepackage{nameref,hyperref}

\usepackage[right]{lineno}

\usepackage{microtype}
\DisableLigatures[f]{encoding = *, family = * }

\raggedright
\usepackage{changepage}

\usepackage[aboveskip=1pt,labelfont=bf,labelsep=period,singlelinecheck=off]{caption}

\makeatletter
\renewcommand{\@biblabel}[1]{\quad#1.}
\makeatother

\usepackage{lastpage,fancyhdr,graphicx}
\usepackage{epstopdf}
\fancyheadoffset[L]{2.25in}
\fancyfootoffset[L]{2.25in}

\usepackage{color}

\definecolor{Gray}{gray}{.25}

\usepackage{graphicx}

\usepackage{sidecap}

\usepackage{tabularx}
\usepackage{mathrsfs}
\usepackage{lineno}
\usepackage{epsfig}
\usepackage{amsmath}
\usepackage{amssymb}
\usepackage{float}
\usepackage{enumerate}
\usepackage{lineno}
\usepackage{booktabs}
\usepackage{arydshln}

\usepackage{wrapfig}
\usepackage[pscoord]{eso-pic}
\usepackage[fulladjust]{marginnote}
\reversemarginpar

\begin{document}
\vspace*{0.35in}

% title goes here:
\begin{flushleft}
{\Large
\textbf\newline{Breakdown modes of capacitively coupled plasma II: unsustainable discharges}
}
\newline
% authors go here:
\\
Hao Wu \textsuperscript{1}
Ran An\textsuperscript{1}
Dong Zhong\textsuperscript{1}
Wei Jiang\textsuperscript{2,*}
Ya Zhang\textsuperscript{3,*}
% Zili Chen \textsuperscript{1},
% Hongyu Wang \textsuperscript{2,*},
% Shimin Yu \textsuperscript{1},
% Yu Wang \textsuperscript{3},
% Zhipeng Chen \textsuperscript{1},
% Wei Jiang \textsuperscript{3},
% Julian Schulze \textsuperscript{4,*}
% Ya Zhang \textsuperscript{5,*}
\\
\bigskip
\bf{1} School of Electronics and Information Engineering, Hubei University of Science and Technology, Xianning 437100, China
\\
\bf{2} School of Physics, Huazhong University of Science and Technology, Wuhan 430074, China
\\
\bf{3} Department of Physics, Wuhan University of Technology, Wuhan 430070,China
\\
\bigskip
* Corresponding Authors: weijiang@hust.edu.cn, yazhang@whut.edu.cn
\end{flushleft}

\section*{Abstract}
In this work, the one-dimensional implicit particle-in-cell/Monte-Carlo collision code (PIC/MCC) is used to study the discharge of a capacitively coupled plasma (CCP) under extremely low pressure driven by high-frequency rf power in pure argon. With the introduction of high-coefficient electron-induced secondary electron emission (ESEE) and a blocking capacitor, the discharge that cannot be sustained shows a variety of different characteristics: including the normal failure discharge (NFD) of the electron avalanche, bias failure discharge caused by the charging effect of the blocking capacitor, and runaway failure discharge caused by the decrease in the ESEE rate during the forming of the sheath. The discharges in low-pressure regions exhibit a range of discharge characteristics, the sustainable discharges of which have been analyzed in more detail. The study of unsustainable discharge helps to find the reasons for failure discharge and then determine the parameters of sustainable discharge, which is of great value in preventing plasma crack, equipment product yield, and equipment safety to help prevent industrial losses.
% now start line numbers
% \linenumbers

% the * after section prevents numbering
\section{introduction}
   Low pressure capacitively coupled plasma driven by rf, when used as a semiconductor process of etching, deposition, or magnetron sputtering, can maintain a long-term stable discharge state in which the particles and power are balanced and most of the physical parameters do not change much over time\cite{lieberman2005principles,chen2023note}. Therefore, most of the research on CCP or capacitive-like devices is focused on the discharge of steady state. The parameters that can break down the gas and sustain the discharge are usually located inside Paschen's curve, which has been the crucial topic of gas discharge\cite{lisovskiy2008similarity,fu2020electrical,jin2022particle,jiang2022gas}.
   Even though failure or unsustainable cases in discharges of CCP or capacitive-like devices are still valuable to investigate:
   
   Almost all of the plasma sources that we used have to undergo the breakdown process. Successful breakdown is a prerequisite for the application of these plasma sources. The study of unsustainable discharge helps to find the reasons for failure discharge and then determine the parameter boundaries of sustainable discharge, which is of great value in preventing plasma crack, equipment product yield, and equipment safety to help avoid industrial losses.
   In addition, some failure and unsustainable discharges are useful for some special processes, e.g. for the high aspect ratio etch processes, where the charging effect will change the anisotropy features of the bombarding ions. Failure discharge will release electrons or negative ions to neutralize the charge in the deeply etched ruts. Unsustainable CCP driven by the pulsed waveform rf that provides intermittent avalanche breakdown and extinction is one of the choices to solve this problem \cite{banna2012pulsed,agarwal2012extraction,adamovich20172017}.
   
   Moreover, in the region of high voltage insulation, on the one hand, gas breakdown should be avoided in some cases to protect the device. On the other hand, once the breakdown occurs, the arc should also be extinguished as quickly as possible to avoid electrical equipment burning or leakage. Therefore, in this region, the dielectric gas is usually used as an insulating material \cite{geng2015experimental,zhang2016experiment,seeger2017breakdown,fu_effect_2017,fu_electrical_2020} to fail gas breakdown or plasma disappearance.

   In high-power microwave devices, multipactor discharge caused by electron-induced secondary electron emission (ESEE) might break the vacuum gap and cause some signal interference and device damage\cite{hohn1997transition,udiljak2003new,wen2022higher}. 
   For those devices, operations and treatments that cause discharge to fail or make it unsustainable will be beneficial for long-working and stable operation to produce microwaves\cite{zhang2019suppression}.
   In the presence of space charge, a high initial seed current density can shrink multipactor susceptibility bands\cite{iqbal2023two}. 

   The failure discharge of CCP at higher pressure (Torr) is relatively simple, in which the mean free path is much lower than the discharge gap. Whether it is driven by DC or RF power, the breakdown voltage will increase with increasing pressure, since collision ionization is dominant \cite{sato1997breakdown}. 
   Most plasma is got in a low background gas pressure of tens to hundreds of millimeters at a lower voltage, in which the glow discharge is easier to start and can be sustained for a long time.

   When the mean free path is lower than the discharge gap, usually in a low pressure (< 100 mTorr), due to the effect of the boundary (electron absorption, reflect and induced SEE), the discharge will show a complex variety of characteristics\cite{lisovskiy1998rf}.
   In discharge driven by rf power, electron-induced SEE (ESEE) plays a significant role in both breakdown\cite{vender1996simulations,lisovskiy1998rf,wu2021electrical}, extinguish\cite{hohn1997transition,lisovskiy2005extinction}, and the stable plasma parameters\cite{horvath2017role,horvath2018effect}. Due to the addition of electrons, the presence of ESEE can significantly reduce the breakdown condition, deviating the breakdown curve toward the low pressure and low voltage region, forming a multivalued Paschen curve\cite{lisovskiy1998rf,wu2021electrical}. 
   
   With a drop in pressure from tens of milligrams to several milligrams and lower, the discharge parameters of the glow discharge reach the limit and will be unsustainable, or change from the glow discharge to multipactor\cite{hohn1997transition,kim2006transition,na2019analysis,hubble2017multipactor,spektor2018space,feldman2018effects}. In high-power microwave regions, window breakdown initiated by the ESEE multipactor discharge is negative for the formation and propagation of electromagnetic waves. So, multipactor discharge, especially plasma formation caused by collision ionization initiated by the multipactor, should be avoided\cite{zhang2019suppression,wen2022higher}.
   Although many excellent works have been applied in the transition of the rf plasma and multipactor through experiments\cite{hohn1997transition} and calculations\cite{kim2006transition,feldman2018effects,zhang2019suppression,wen2022higher,iqbal2023recent}.
   
   Some questions remain to be answered, especially about the evolution of the key parameter of failure discharge in the transition region between glow discharge and multipactor discharge. 
   With the help of the particle-in-cell / Monte Carlo collision (PIC / MC) code coupled with a comprehensive ESEE model\cite{horvath2017role}, the transition of failure discharge at extremely low pressure driven by a high frequency of rf power will be investigated in more detail.
   As the second part in the series, this paper is organized as follows: the methods used in this work will be introduced in Section 2. The simulation results of different unsustainable discharge modes will be presented and analyzed in Section 3. Furthermore, the unsustainable reason will be discussed in Section 4. In the end, the conclusions will be drawn in Section 5.
  
\section{Methods}

  %External circuit\\
  Two flat electrodes $0.031415926m^2$ are placed symmetrically and opposite to each other to form the discharge gap, and their distance is set to 0.02 m. A 200 pF blocking capacitor is connected between the powered electrode (electrode1) and the rf power source, and another electrode(electrode2) is grounded. The RF power waveform satisfies $U_S(t) = U\sin{(2 \pi ft)}$, where $U$ and $f$ represent the voltage amplitude and frequency, respectively. $f$ is set to 60MHz.

  %see proposed by Horváth in 2017
  % Both the experiments and simulation results show that the SEE induced by ions in the breakdown process can be ignored. However, the SEE induced by electrons plays an important role in rf-driven breakdown\cite{smith2003breakdown,vender1996simulations,radmilovic2005modeling}, which can significantly reduce the minimum breakdown pressure in low pressure region of Paschen's Curve.
  There will be a lot of electrons bombarding the electrodes during the rf driven discharge. If the energy of the primary electron is low, it might be absorbed or reflected. If the energy is greater than the emission threshold of the electrode surface (a dozen electron volts), the electron might be reflected, or absorbed, and a new secondary electron might be induced and emitted from the surface of the electrode to the gap. Therefore, the coefficient of ESEE ($\delta$) should be a function of the incident electron energy and is affected by the type of electrode material and surface roughness\cite{verboncoeur2005particle,horvath2017role,horvath2018effect}. 
  In this work, the ESEE mode summarized by Horvath\cite{horvath2017role, horvath2018effect} is used to depict the secondary electron emission in this work, which is suitable for flat SiO$_2$ electrodes. The real SEE, elastically reflected electrons, and inelastically reflected electrons have been treated separately in Horvath's work.
  In aluminum electrodes, the aluminum oxide film on the surface will make the emission coefficient greater than that of SiO$_2$ \cite{guo2019secondary}. In fact, in the high incident electron energy region (several hundred electron volts), the trend of the emission function is similar in different SEE modes \cite{vender1996simulations,smith2003breakdown}. In this work, $\sigma_{max}$
  is used to control the ESEE coefficient. 
  %code
  PIC/MCC has been widely used to simulate glow discharge and multipactor \cite{kim2006transition,na2019analysis}.
  In this work, we used one-dimensional direct implicit particle-in-cell/Monte Carlo collision (PIC/MCC) coupled with an external circuit. 
  The time step and the grid number are set to $5.0\times 10^-11$ and 65 respectively. 
  Since the implicit code allows for a larger time and space scale than the explicit\cite{vahedi1993capacitive,kawamura2000physical,wang2010implicit}, this can finish the simulation more quickly.

   Pure argon is used as the background gas, and only the electron and Ar$^+$ are traced in this simulation. The consumption of the background gas is ignored since the ionization rate is extremely low in rf-driven low-pressure discharge.
   The standard MCC model proposed by K.Nanbu and Vahedi is adopted to deal with collisions \cite{vahedi1995monte}. The cross-section data come from\cite{phelps1999cold}. For electron-argon (e-Ar) collisions, elastic scattering, excitation, and ionization collisions are considered. For ion-argon(e-Ar$^+$) collision, only charge exchange and elastic scattering are considered. 

   In this work, two simulating methods will be used. To explore the fast formation of a glow discharge or multipactor, the whole discharge process from the extremely low electron density ($10^{8}m^{-3}$) to the last stable discharge state will be "diagnosed" and drawn. The final stable discharge is often the working state in most plasma sources, and it is also the state that can be relatively easy diagnosed through experiments. To better know the steady state of discharge, more diagnostic codes will be added and run in more repeated periods to obtain the convergence result as accurately as possible.
%    %给出
\section{Results}
  \subsection{breakdown curve}
\begin{figure}[ht]
       \centering
         \includegraphics[width=0.7\textwidth]{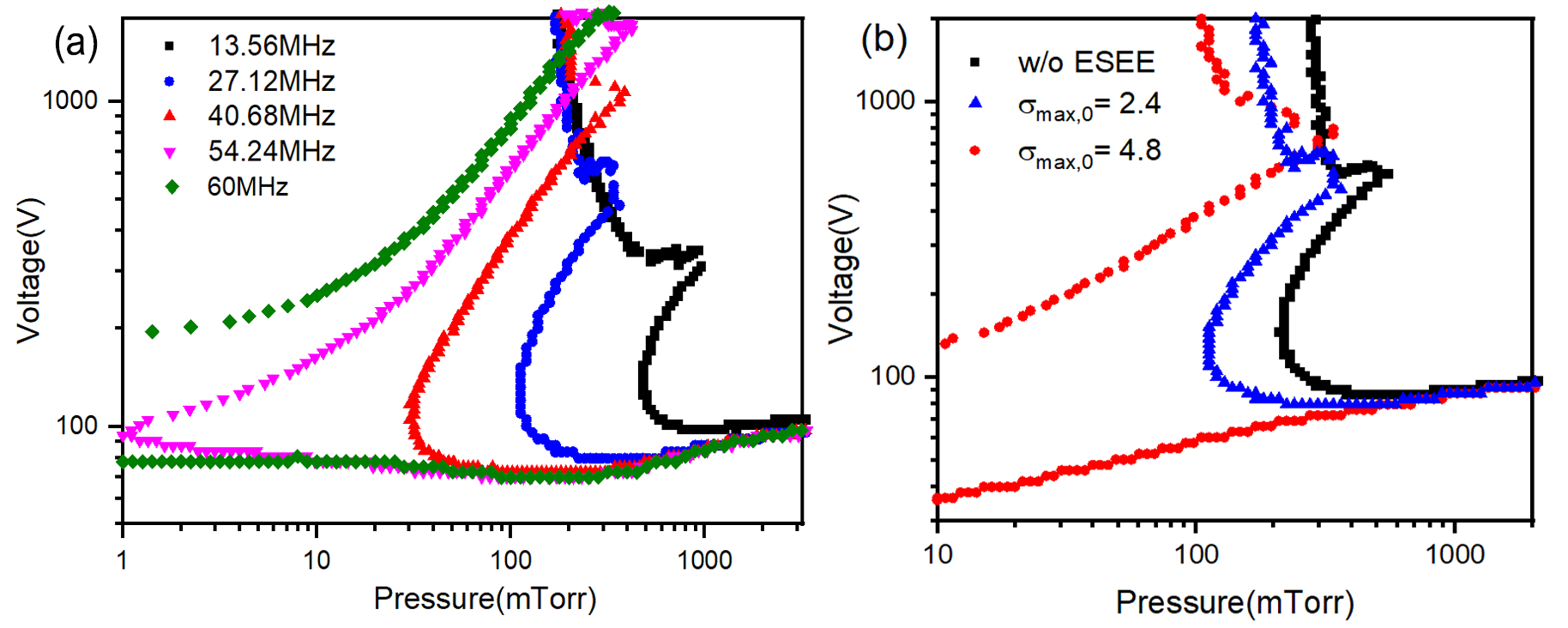}
         \caption{Breakdown curve argon under different discharge conditions: (a) different rf frequency ($\sigma_{max} = 2.4$), (b) different $\sigma_{max}$ under 27.12 MHz. The dots denote the threshold boundary if an electron avalanche can occur.}
      \label{BreakdownCurve}
    \end{figure}

     Through PIC/MC code, starting from a low pressure, gradually increase the pressure until an electronic avalanche occurs, or start from a low voltage, gradually increase the voltage amplitude and record the pressure voltage threshold at which an electronic avalanche occurs. 
     After scanning the pressure-voltage (V-P) zone, the breakdown curve is obtained under different circumstances, as shown in figure \ref{BreakdownCurve}. 
      To know whether the gas can be broken down by judging whether the avalanche occurs is a theoretical method for obtaining a theoretical Paschen curve in the previous discharge model, 
     such as the Townsend model \cite{townsend1915electricity,lieberman2005principles} and Monte-Carlo simulation \cite{korolov2014experimental,puavc2018monte,puavc2020monte}.
  
    The breakdown curve driven under low frequency and low ESEE coefficient conditions, as shown in figure \ref{BreakdownCurve}, is similar to the curve obtained by experiment\cite{lisovskiy1998rf} or Monte Carlo simulation\cite{korolov2014experimental}. However, under $\sigma_{max} = 2.4$, with increasing frequency of the rf, the turning point of pressure ($P_{t}$) and voltage ($U_{t}$) will move left until it disappears at growing 60 MHz. The breakdown curve splits from one to two, which means that only the voltage inside the two curves can avalanche take place. Under the higher ESEE $\sigma_{max} = 4.8$, only the 27.12MHz rf power can split the breakdown curve from one to two, as shown in Figure \ref{BreakdownCurve}(b).
   Although the condition of $\sigma_{max} = 4.8$ is difficult to meet. Although the aluminum electrode with aluminum oxide film can reach a value $\sigma_{max}$ of more than 4.0 \cite{guo2019secondary}.
    However, $\sigma_{max} = 2.4$, is relatively easy to reach; most $\sigma_{max}$ is relatively easy to achieve a surface emissivity of 2.4, which can be achieved by many metal oxide surfaces and SiO$_2$ surfaces\cite{vender1996simulations,horvath2017role,smith2003breakdown}. Thus, for most discharges driven at a higher frequency of 60 MHz, one breakdown curve split into two curves in the low-pressure region will easily occur.
    This feature was analyzed many years ago \cite{hohn1997transition} in the transition of glow discharge to multipactor. The discharges in low-pressure regions exhibit a variety of discharge characteristics, of which the sustainable discharges driven by 60 MHz have been analyzed in Part. I. In fact, there are many types of failure or unsustainable discharges inside the two breakdown curves that deserve attention. In this part, failure or unsustainable discharges under the frequency of 60 MHz will be given and analyzed.

   \subsection{Discharge mode inter two breakdown curve}
     After a wide range of voltage and pressure scanning, it is found that there are several discharge modes that the discharge cannot be sustained, which has been shown in square regions of figure 1 in Part I. That means the particle density will go down or even disappear; the cyan triangle region is the unstable transition zone, the discharge will not collapse but cannot be stabilized. 
%     % \begin{figure}[ht]
%     %    \centering
%     %      \includegraphics[width=0.7\textwidth]{PSST20230206/PartIFigures/DischargeScanning.png}
%     %      \caption{The distribution of Different discharge modes in the frequency of 60MHz and distance of 2cm.}
%     %   \label{BreakdownCondition}
%     % \end{figure}
    Just like Part I, to better describe discharges, we also renamed each unsustainable mode and gave the representative points (displacement parameters with representative discharge characteristics) of them shown in the table \ref{unSustainableDischargeName}. 
   Three types of unsustainable discharge can be divided into normal failure discharge (NFD), runaway failure discharge (RFD), and bias failure discharge(BFD). 
   
  \begin{table}[ht]
        \centering
        \caption{Different unsustainable discharge modes in 60MHz 2cm}
        \begin{tabular}{ccc}\hline
           Discharge mode  &  Abbreviate   &  Representative Point\\\hline
           normal failure discharge   &  NFD  & 75 V 0.5 mTorr \\
           runaway failure discharge &  RFD   & 160 V 2.0 mTorr \\
           bias failure discharge & BFD & 90 V 0.5 mTorr \\
        %   Unsustainable discharge & UD &  70V 0.5mTorr  \\
           \hline
        \end{tabular}
        \label{unSustainableDischargeName}
    \end{table}

\begin{figure}[ht]
       \centering
         \includegraphics[width=0.5\textwidth]{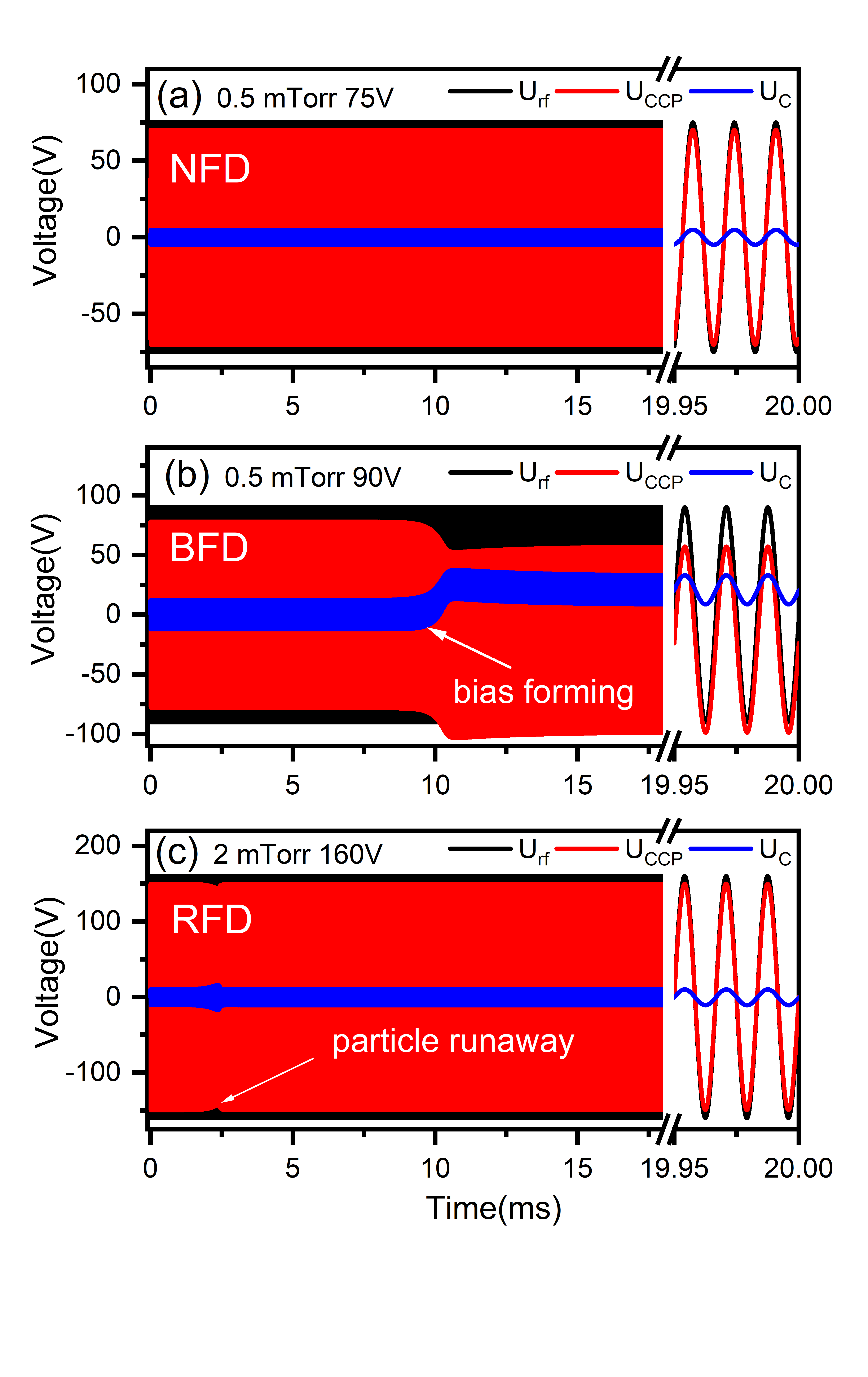}
         \caption{Evolution of the circuit signal, (a) NFD (0.5 mTorr, 75V), (b) BFD (0.5 mTorr, 90V), (c) RFD (2 mTorr, 160V)}
      \label{UnstableECUsUccpUc}
    \end{figure}

     From the external circuit electrical signal evolution in figure \ref{UnstableECUsUccpUc}, and the electron density evolution, three different failure dischargees can be well distinguished.
    
  \subsection{NFD -- normal failure discharge}
    The normal one is the traditional failure due to insufficient electron yield under certain conditions, that is, the electron avalanche cannot happen.  
    The electron density of NFD has never increased under the average radio frequency cycle, and has been declining since the beginning of discharge. Because of the extremely low electron density, the discharge chamber almost has no effect on the external circuit, resulting stable voltage waveform of figure \ref{UnstableECUsUccpUc}(a).
    
\begin{figure}[ht]
       \centering
         \includegraphics[width=\textwidth]{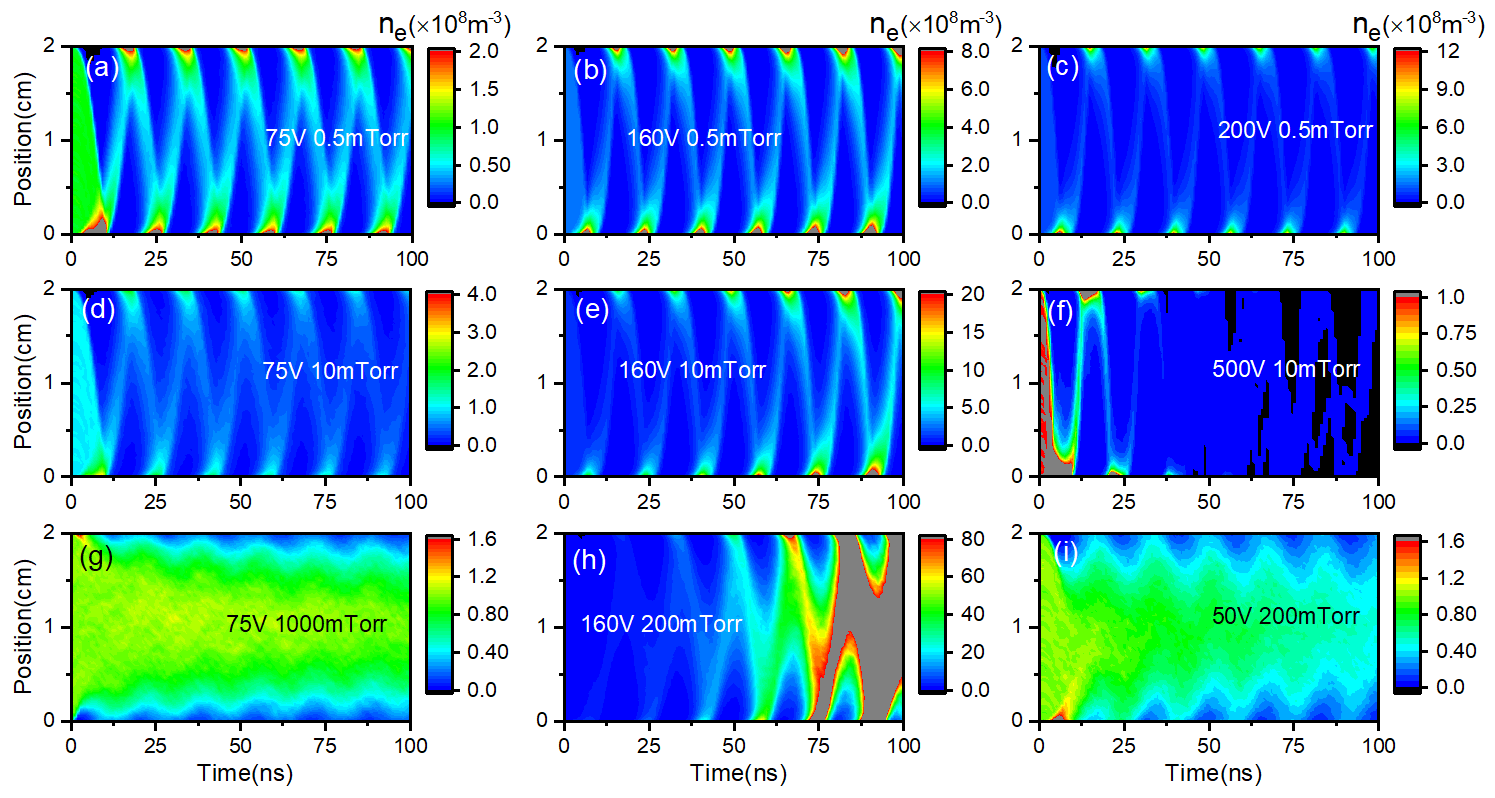}
         \caption{Spatio and initial-temporal evolution of electron density in different discharges model under the frequency of 60 MHz: (b),(e), and (h) are the modes that can be successfully broken down; (a), (d), (g), and (i) is the NFD of insufficient voltage; (c), (f) are the NFD of overhigh voltage}
      \label{NormalFailureNe}
    \end{figure}

    NFDs under different conditions are listed in figure \ref{NormalFailureNe}. To better represent them, the multipactor discharge under extremely low pressure of 0.5 mTorr, the glow discharge induced by ESEE and ionization collision have also been listed, as shown in Figure \ref{NormalFailureNe} (b), (e), and (h), respectively.
    
    In the traditional unsustainable discharge of CCP, NFD is located in the outside region of the Pachen curve. The breakdown curve presents a single value characteristic in the higher pressure region. Ionization within the discharge gap is the main way to generate a new electron-ion pair, as shown in Figure \ref{NormalFailureNe}(g, h, i). The requirement for electrons to obtain the energy required for ionization collisions on a mean-free path is a necessary condition for successful breakdown. 
    In higher pressures of 200 mTorr or 1000 mTorr, a lower voltage cannot result in electron avalanche, and the electron density decreases directly caused by the insufficient electron energy, as shown in Figs. \ref{NormalFailureNe}(g) and \ref{NormalFailureNe}(i). 
     Thus, since the pressure is inversely proportional to the mean free path, in the case of high gas pressure, the breakdown voltage and gas pressure show a relatively good linear characteristic \cite{lisovskiy1998rf}. Note that the high pressure here refers to the fixed discharge gap, and the linear region will move to the higher pressure when the discharge gap becomes smaller.
     
     When the pressure of background gas declines to 200 mTorr and lower, electrons have more chances to bombard the boundary, at this time, the boundary effect gradually appears, and the disappearance of plenty of electrons will fail the breakdown. Therefore, when the gas pressure drops to 100 mTorr or less, the minimum breakdown voltage does not decrease with decreasing pressure but instead increases.
     
     The rf field gives the electrons the opportunity to change their direction of motion, so that rf-driven spectroscopy is more accessible to break down the gas than a DC.
     However, a higher voltage still has more chance to push the electrons that bombard the electrode. 
     If the voltage is high enough, the electrons will have no chance of turning around and disappearing on the electrode. Therefore, driven under the rf power, the breakdown curve presents noticeable multivalued features in the lower pressure for the effect of boundary loss.
     When the SEE module is not considered, the minimum pressure of the Paschen curve is 100 mTorr, just as shown in the red line of Figure 1 of Part I. If the frequency is high enough, the moving electrons will be diverted before hitting the electrodes and gain energy to continue ionizing the gas. Therefore, the high-frequency rf-driven breakdown curve has apparent multivalued features in the low-voltage region. 
   
     The existence of ESEE significantly expands the low pressure range that can be broken down, as shown in the green circle region of Figure 1 of Part I, which is the same as previously concluded\cite{smith2003breakdown,radmilovic2005modeling}.
     %It should be noticed, only the breakdown voltage within a specific range in the lower-pressure region is broadened when the ESEE is considered.
    The effect of ESEE is more significant only when the boundary effect is significant (lower gas pressure or smaller spacing), which is mainly affected by three cases: (1) the secondary electrons bombarding the electrode should have a high enough energy to reach the electron emission threshold of the electrode surface. This requires that the rf voltage be higher than a value; (2) The newly emitted electrons should have enough opportunity to be accelerated to the center of the discharge gap by the electric field rather than being held back by the original electric field to the surface of the electrode.
    This requires that the amplitude of rf voltage is not too high, or the electrons just emitted will be dragged back to the electrode;
    (3) a lower frequency will slow the change period of field direction, and the newly emitted electrons will be easily pulled back, so the high frequency makes it easier to retain the newly emitted secondary electrons. If the electrons arrive before the electric field changes its polarity, the secondary electrons emitted are repelled\cite{hohn1997transition}. Ideally, when a large number of electrons bombard one electrode, the direction of the electric field just changes, and the role of the SEEs is most significant at this time.
    
    Therefore, the discharge dominated by ESEEs will make the multi-value of the breakdown voltage more obvious. In the range of 5$\sim$200 mTorr, both the lower or higher voltage will fail the breakdown, as shown in Figure 1 of Part I. Thus, at low pressure of 10 mTorr, only in the voltage range of 75$\sim$200V can create the glow discharge, and the higher or lower rf voltage will lead to discharge failure, as shown in Figures \ref{NormalFailureNe}(d), (e) and (f). In this case, the electron density near the electrode is significantly higher in the pre-breakdown phase, indicating that ESEE almost dominates the electron avalanche.

    When the gas pressure drops to a pressure lower than 1 mTorr, the ionization collisions in the gap are almost negligible relative to the secondary electron emission. 
    In this field, only the interaction between charge and electric field is prominent, that is, single-particle orbital theory. In this case, the energy of the electrons bombarding the electrode can only be affected by the rf voltage, frequency, and distance. For the parameters of this work (60 MHz, 2cm), only in a certain voltage range (80$\sim$180V), the ESEE-dominated electron avalanche can happen, that is, electron multipactor discharge, just as shown in figure \ref{NormalFailureNe}(b). This voltage range hardly changes with changes in gas pressure when the pressure is lower than 2 mTorr. When the voltage is lower than 80V, the electrons in the gap cannot obtain enough energy to generate more secondary electrons, so the electron density gradually decreases in the initial stage of discharge, as shown in Figure \ref{NormalFailureNe}(a). 
    When the voltage is higher than 160V, the newly emitted electrons will be pulled back by the original strong electric field and lost on the surface of the electrode surface, which also causes the decrease in electron density, as shown in figure \ref{NormalFailureNe}(c). 
    It should be noted that even though the electron avalanche can occur in the pressure range of $p < 2$ mTorr and the voltage range of 80$\sim$180V. There will still be other reasons for the electrons in the gap to disappear, causing the discharge to fail, which will be introduced in \ref{RunawaySection} and \ref{BiasSection}.

   \subsection{RFD -- runaway failure discharge} \label{RunawaySection}
      %160 V 2.0 mTorr 有无种子电子的密度电势图
      Discharge characteristics in transitional regions are always worthy of attention. At a voltage of 120$\sim$180 V, the glow discharge will turn to a multipactor when the gas pressure drops from 5 mTorr to 0.5 mTorr. 
      There will be a narrow transition region around 2mTorr, in which the electron avalanche can occur but the discharge cannot be sustained, just as shown in the magenta square zone of the figure 1 of Part I.
      
      \begin{figure}[ht]
       \centering
         \includegraphics[width=\textwidth]{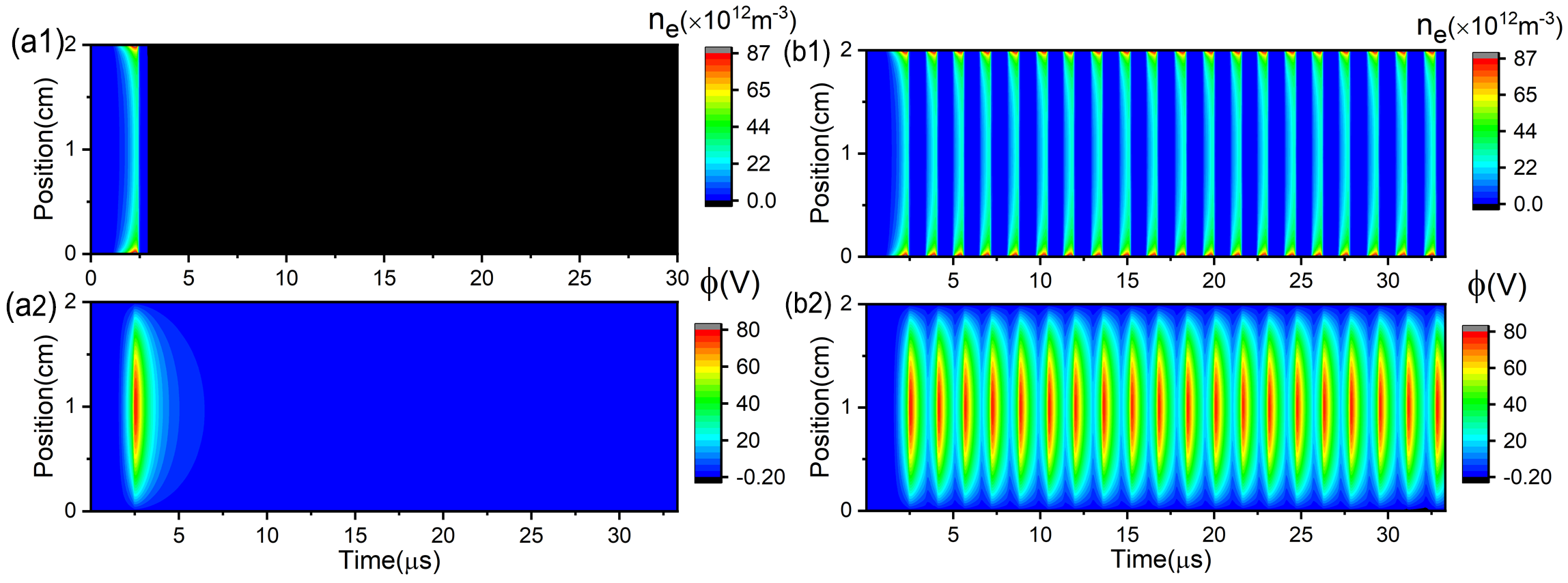}
         \caption{RFD:(a1),(a2) is the spatio-temporal evolution of electron density and potential under the existence of 200 pF blocking capacitor; (b1), (b2) is the results after introducing the seed electron. 60MHz 160V 2mTorr}
      \label{RunawayFailureNePhi}
    \end{figure}

    When the pressure drops to 2 mTorr, a failure discharge mode is obtained in the transition region between the discharge region of glow discharge and the multipactor. The electron density of this transition discharge mode are shown in figure \ref{RunawayFailureNePhi}(a1),(a2).
    After 2.5 microseconds of electron avalanche, the electron density is close to $\times 10^13$m$^{-3}$. The self-generated potential of more than 80 V is observed in the center under the applied voltage of 160 V, which means the plasma begins to form.
    However, for some reason, the electron density suddenly drops to zero, and the potential shifts from a sheath-like distribution to a diffuse field distribution, as shown in figures\ref{RunawayFailureNePhi}(a1) and (a2).
    Due to the escape of electrons, the breakdown cannot be completed, so we named this type of discharge runaway failure discharge.
    Because of the escape of electron, the formation of the sheath stops abruptly, thus causing the DC-blocking capacitor voltage amplitude to immediately return to its initial value after a brief rise, as shown in figure \ref{UnstableECUsUccpUc}(c).

    What is interesting is the pulse-driven discharge, which is obtained after the seed electron is introduced immediately when the electron density declines to zero. In fact, in the real environment, not all electrons can escape, and there will always be seed electrons inside the discharge through cosmic background ionization or field emission from the electrode surface, etc.
    Therefore, figure\ref{RunawayFailureNePhi}(b1) and (b2) will be the discharge mode that we observed in real discharge instead of figure\ref{RunawayFailureNePhi}(a1) and (a2). This spontaneously formed periodicity will periodically release electrons with an energy of more than 20 eV to the electrode, which is helpful for some of the high aspect ratio dielectric etching processes in the semiconductor industry.

\begin{figure}[ht]
       \centering
         \includegraphics[width=0.8\textwidth]{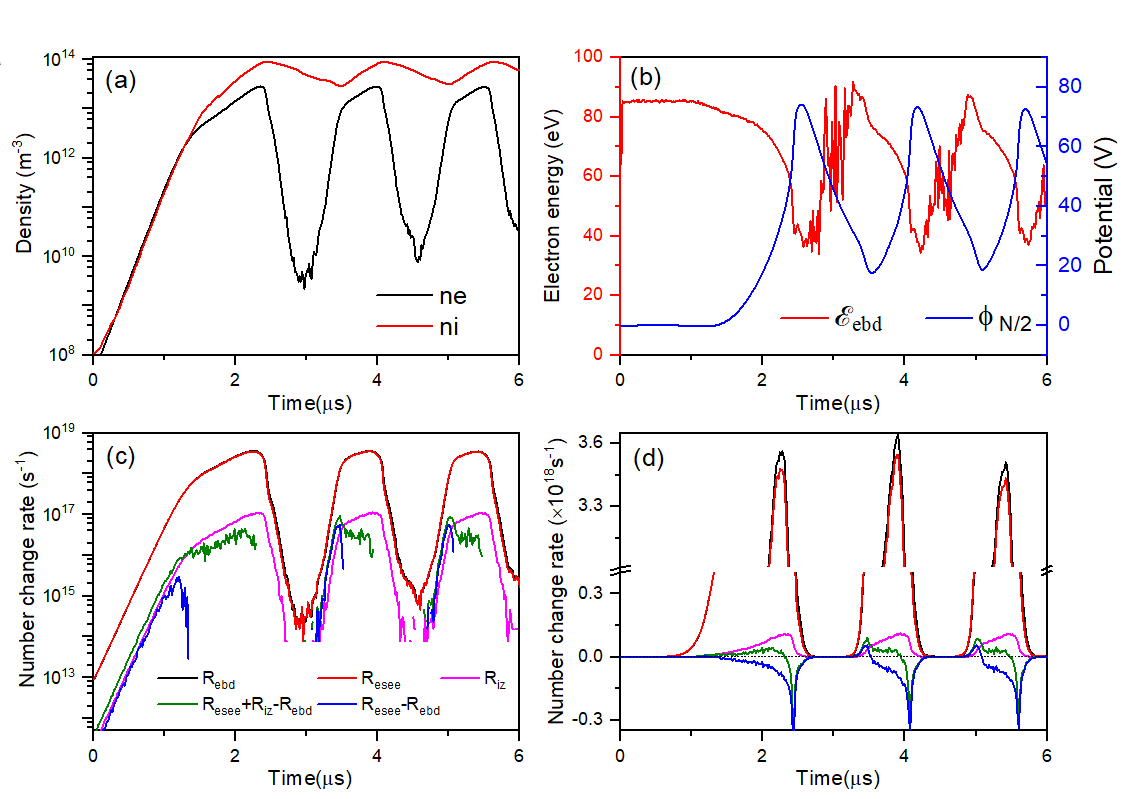}
         \caption{Key parameter of RFD (60MHz, 160V 2.0mTorr): (a) center electron and ion density, (b) mean electron bombarding the electrode(red line) and center potential, (c) different terms of electron number change rate in logarithmic coordinates ($R_{ebd}$ indicates the electron flux through electrodes $R_{esee}$ is the electron generation rate caused by ESEE, $R_{iz}$ denotes the electron generation rate caused by ionization), (d) linear ordinate of (c)}
      \label{RunawayParticleBalance}
    \end{figure}

    %分析该放电产生的原因
    Compared to the discharge of 10 mTorr (low-pressure glow discharge), the mean free path of electrons has increased by nearly an order of magnitude, which directly caused the insufficient ionization rate.
    %When gas pressure drops to about 2 mTorr, compared to the NSD discharge above 5 mTorr, the ionization rate will be much lower than the SEE emission rate due to the lower gas pressure. 
    When the electron density rises to a critical value (about 5$\times 10^12$m$^{-3}$), the electrons near the electrode will have more chance to escape to the electrode quickly, and the sheath begins to form. %the self-generated field  
    Due to the high mass of positive ions, the central potential rapidly increases above 50 V. %still trapped in the gap
    The rapid growth of the positive potential of the core greatly reduces the electron energy bombarded to the electrode (shown in figure \ref{RunawayParticleBalance} (b)), reducing the probability of SEE emission of a single electron. 
    At this time, the sheath is usually one-sided and unstable, oscillating greatly with the applied electric field. 
    This causes the electrons to move on the other side and escape quickly and continues to increase the center potential. When the mean probability of SEE emission drops to a threshold value, the SEE emission rate ($R_{esee}$) and the ionization rate ($R_{iz}$) will be lower than the electron loss rate at the boundary ($R_{ebd}$), as shown in figures \ref{RunawayParticleBalance} (c) and (d). 
    This will cause the electron density to decrease rather than increase as shown in the black line of figure \ref{RunawayParticleBalance}(a), so that the newly unstable sheath becomes more fragile, causing electrons to continue to disappear from the boundary rapidly until most of the electrons escape.
    This discharge caused by electron escape is named runaway failure discharge (RFD) in this work, which is located in the area of the magenta square symbol in figure 1 of Part I, in the gas pressure range of 1$\sim$ 5 mTorr and the voltage range of 80$\sim$180 V.

    This discharge failure is abnormal, as electron avalanches have occurred in the initial phase, during which electrons and ions continuously accumulate.
    Extensive particle escape occurs during sheath formation, leading to a sudden collapse of the newly formed unstable sheath. 

    The introduction of the seed electron after the electron density declines to zero since the outside discharge parameter is not changed, the electron avalanche will occur again, causing the electron density to increase again, and the electrons after reaching a certain critical value will escape again, forming a microsecond-level "flicker" discharge. It should be noticed that most of the ions generated by previous discharge still exist in the gap for their much higher mass; more than 30 V center positive potential still exists when the electron declines to zero. The 30 V central potential will play a certain binding role on the electrons generated during the second electron avalanche process. Therefore, the second and third avalanches will be faster than the first process. %The 30 V central potential will play a certain binding role on the electrons generated during the second electron avalanche process, making the electron density increase faster in the subsequent discharge.
       
  \subsection{BFD -- bias failure discharge} \label{BiasSection}
     %90 V 0.5 mTorr 电子密度电势图
   In most capacitive discharges, there will be a blocking capacitor connected between the rf power and the discharge device. Thus, compared to the simulation result without an external circuit, the simulation results in the lower voltage region are almost opposite
    
\begin{figure}[ht]
       \centering
         \includegraphics[width=0.8\textwidth]{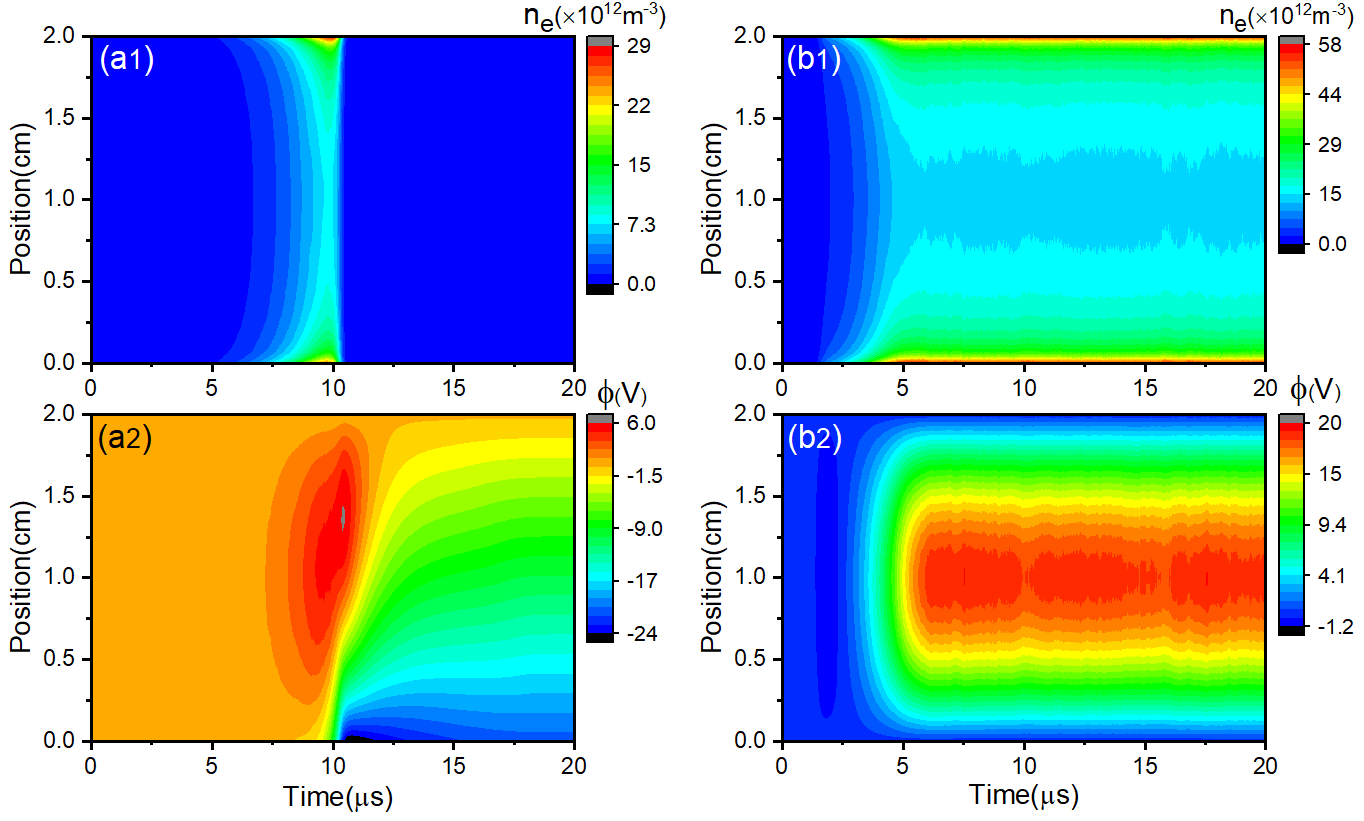}
         \caption{Key parameter of BFD(60MHz 90V 0.5mTorr):(a1),(a2) is the spatio-temporal evolution of electron density and potential under the exist of 200 pF blocking capacitor; (b1) is the results without blocking capacitor.}
      \label{BiasFailureNePhi}
    \end{figure}

   Between two regions of NFD under extremely low pressure (lower than 1 mTorr), there will be a multipactor between the voltage of 80$\sim$ 180V at a frequency of 60 MHz. When the dc blocking capacitor is not considered, the discharge does exhibit a multiplied discharge characteristic between the voltage of 80$\sim$180V, as shown in Figure \ref{BiasFailureNePhi}(b1).
   After considering the blocking capacitor, at the voltage of 120$\sim$ 180V, the discharge type is in fact the multipactor. However, the discharge cannot be sustained at a voltage of 80$\sim$120V, in which the electron density directly decreases from about $3.0\times 10^{13}$ m$^{-3}$ to zero, as shown in Figure \ref{BiasFailureNePhi} (a1). 

    %100 V 0.05 mTorr 电子密度电势图   负偏压失败图
    From the time-spatial distribution of potential in the discharge simulation with and without blocking of figure \ref{BiasFailureNePhi}(a2) and (b2), we can know that an obvious bias potential appeared in the voltage of 80$\sim$120V, which is just like \ref{BiasFailureNePhi}(a2).

    Due to the introduction of dc blocking capacitors, when the gas pressure is lower than 1 mTorr, the avalanche process can still occur in a lower voltage range (85$\sim$120 V). However, as the electron density gradually increases, since a blocking capacitor is connected in series between the CCP and the rf power supply, the imbalance between the electron absorption on the powered electrode and the ESEE emission will charge the blocking capacitor to create bias voltage.
    The formation of bias voltage is also irreversibly reflected on the DC blocking capacitor of the external circuit, as shown in \ref{UnstableECUsUccpUc}(b).

    The self-bias comes in two ways: On the one hand, electrons disappearance on the electrode will charge the blocking capacitor to create negative bias voltage. On the other hand, electrons with higher energy bombarding the electrode will have more chance to create more than one SEE, which will cause the positive bias. It should be noted that both the two bias feedback mechanisms are positive. The negative bias created by the accumulation of electrons will reduce the energy of electrons bombarding the powered electrode and decline the emission coefficient of SEE so that negative charges continue to accumulate. The positive bias created by the higher emission SEE coefficient will increase the energy of those electrons bombarding the electrode, which will increase the SEE emission coefficient and increase the positive bias voltage. 
    
    So this kind of discharge is called the bias failure discharge (BFD) in this work.
    When the applied rf voltage is much higher than the bias voltage, the effect of the bias can be ignored.
    The formation process of weak bias voltage will be covered by strong rf voltage, and the SEE coefficient will not change much, forming a negative feedback mechanism of pure charge and bias voltage, so when the voltage is large (higher than 120 V), the BFD will not occur.
    Therefore, due to the low voltage of 85$\sim$120 V, this bias will form positive feedback, breaking the balance of particle flux at the two electrodes, pushing electrons to the electrode, causing the failure fo the discharge, forming a low voltage red square region of figure 1 of Part I.
    The BFD is completely caused by the DC blocking capacitor. If there is no blocking capacitor, the discharge will become multipactor, as shown in figure \ref{BiasFailureNePhi}(b1) and (b2).
    \begin{figure}[ht]
       \centering
         \includegraphics[width=0.45\textwidth]{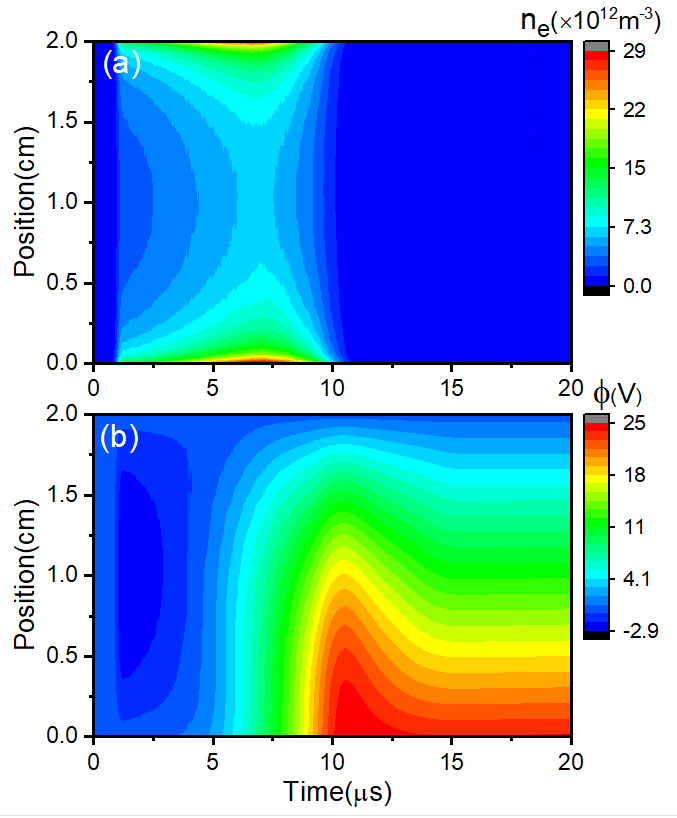}
         \caption{Key parameter of positive BFD (60MHz 100V 0.1mTorr): (a) the spatio-temporal evolution of electron density; (b) the spatio-temporal evolution of potential}
      \label{BiasFailureNePhiPositive}
    \end{figure}
    
    Also, it needs to be noticed that the bias type is uncertain. In most cases, the bias is negative since the electron charging effect of electrons often dominates. We also found several cases of discharge failure caused by positive bias in the BFD region, as shown in figure \ref{BiasFailureNePhiPositive}.

      \begin{figure}[ht]
       \centering
         \includegraphics[width=0.5\textwidth]{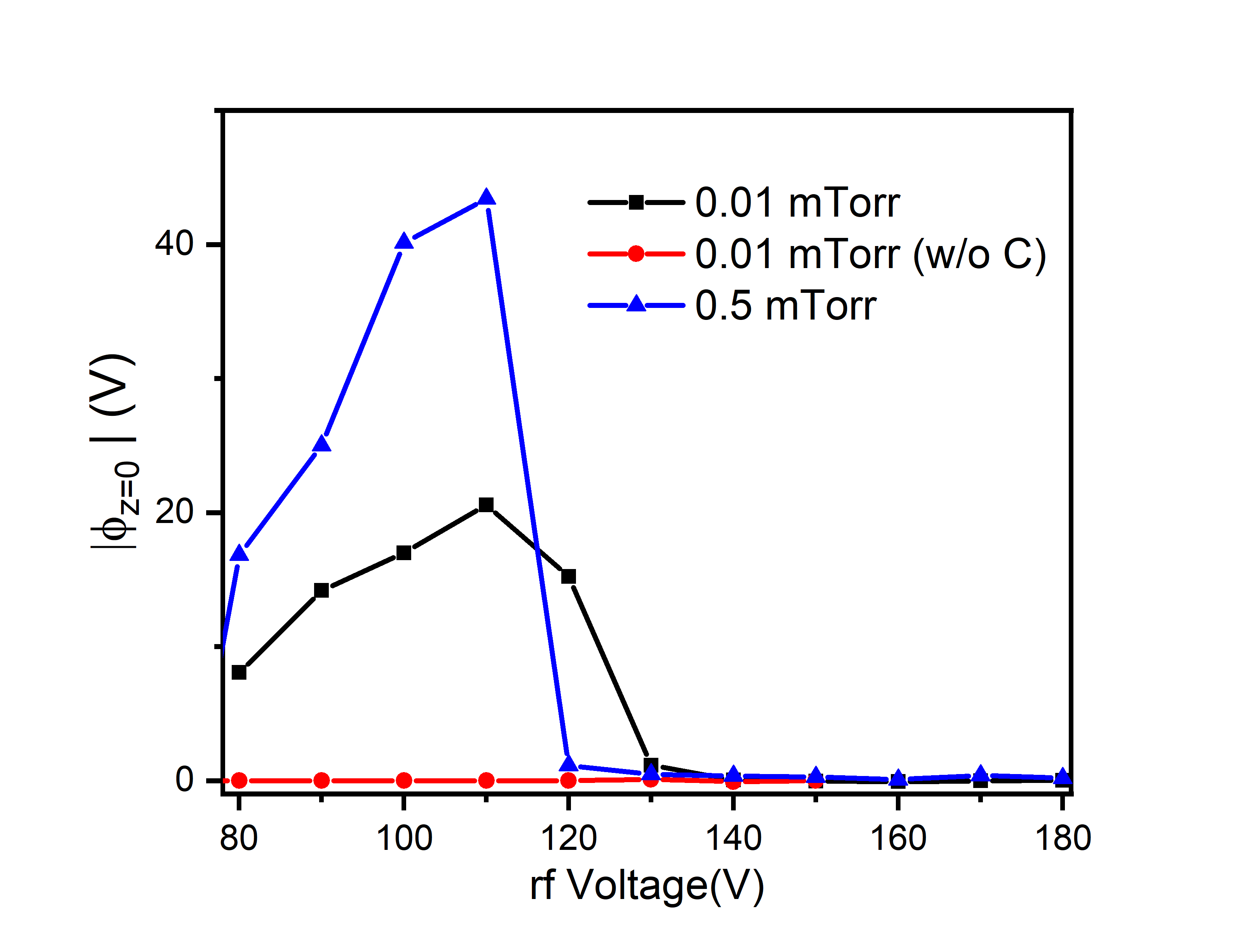}
         \caption{Potential of powered electrode under different voltage of rf power, "w/o C" means the results without the 200 pF blocking capacitor}
      \label{BiasDiffVoltage}
    \end{figure}

 \section{Discussion of the RFD}

\begin{figure}[ht]
       \centering
         \includegraphics[width=\textwidth]{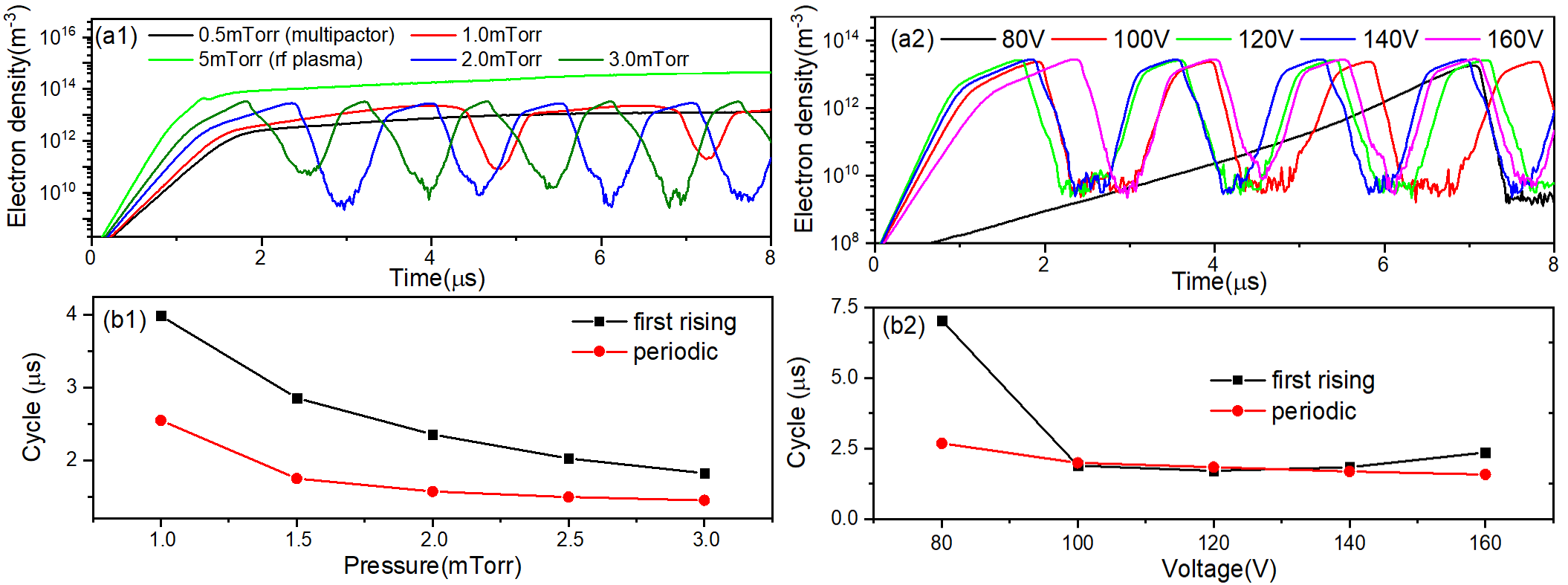}
         \caption{The effect of pressure (voltage fixed to 160 V) and voltage (pressure fixed to 2 mTorr) on the discharge of RFD: (a1) and (a2) is the evolution of center electron density, (b1) and (b2) is the different characteristic times, in which "first rising" means the time of the electron density to rise from the initial density to the density maximum for the first time, "periodic" denotes the mean cycle of the periodic avalanche}
      \label{RFDDiffPVNeCycle}
    \end{figure}

  Although the type of RFD cannot be sustained, the periodic avalanche and sheath forming will make the statistical macroscopic physical parameters full of significance. Self-generated pulse-like discharge might expand the application field of gas discharge. Therefore, how voltage and gas pressure affect the characteristics of RFD is of great significance for discussion. Since the formation process of RFD has been discussed in Section 3.4, this section will focus on the effects of gas pressure and voltage on RFD. 
  The evolution of electron density under different pressures and voltage are shown in figure \ref{RFDDiffPVNeCycle}(a1) and (a2).
  The formation of the self-generated field limits the emission of secondary electrons, causing electrons to run away. Thus, the maximum electron density of RFD will be much lower than glow discharge and higher than multipactor, as shown in figure \ref{RFDDiffPVNeCycle}(a1).
  
  In extremely low gas pressure regions, increasing the gas pressure will directly shorten the free-pass of electrons, which will directly increase the ionization rate. Therefore, the increase in gas pressure can significantly speed up the electron avalanche process, shorten the electron density increasing time, and thus shorten the discontinuity period, just as shown in figures \ref{RFDDiffPVNeCycle}(a1) and (b1).
  Note that because the ions are too heavy, it is almost impossible for the ions to run away during the electron-escaping process. As a result of the retention of ions, there will be a positive potential inside the discharge gap after the electrons run away, forming a potential barrier for the electrons, causing the density of the electrons to grow faster during the second avalanche.

    When the voltage is too high or too low, RFD cannot be formed. In the first avalanche of electron density rising, 120V of the rf voltage amplitude has the fastest density rising speed. However, after the first electron runaway process, the interruption period of discharge gradually decreases as the voltage increases. In the second or third avalanche, under different voltages, there is not much difference in the time required for electron runaway and avalanche. The introduced seed electrons need a certain time to get enough energy to re-examine the electron avalanche process. Therefore, low voltage will have longer periods of low-density basin. This is the main reason that the cycle of the periodic RFD declines with the increase of the rf voltage, as shown in Figure \ref{RFDDiffPVNeCycle} (b2).

    The interaction between charged particles and the electrode surface deserves attention, which is of great significance for expanding plasma applications and limiting its adverse effects.
    \begin{figure}[ht]
       \centering
         \includegraphics[width=\textwidth]{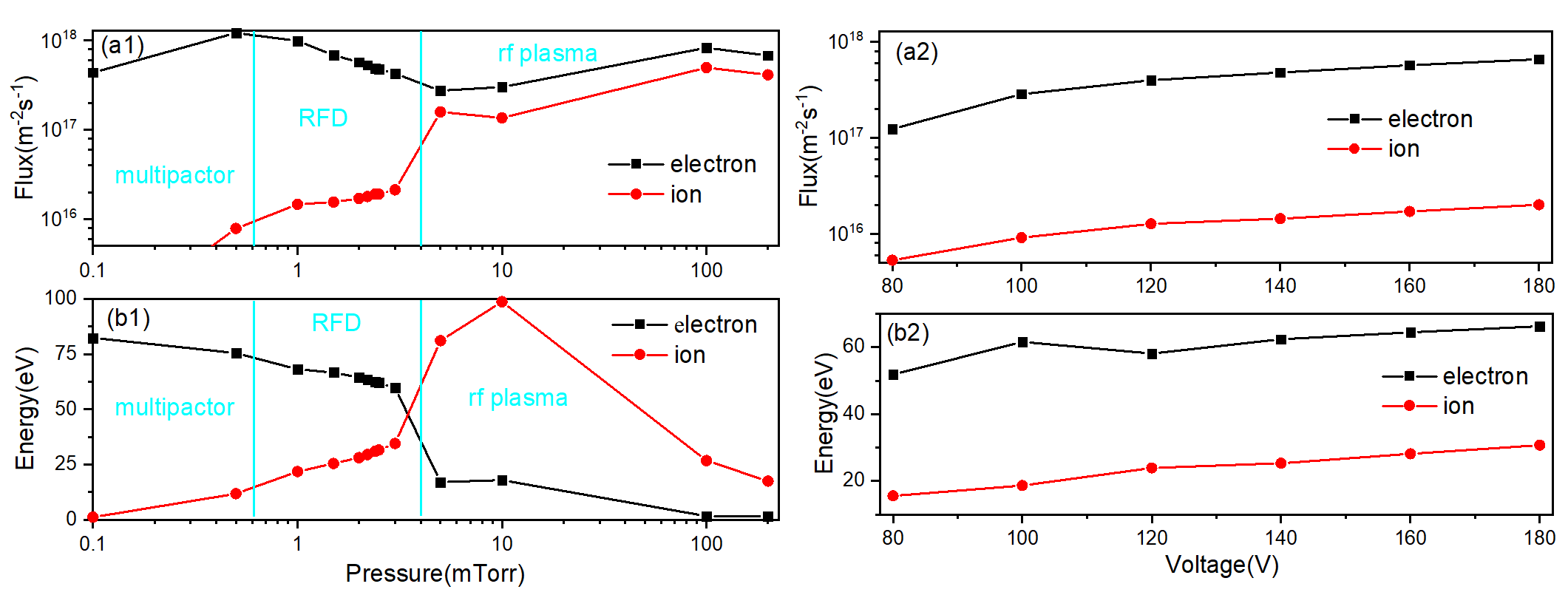}
         \caption{The effect of pressure (voltage fixed to 160 V) and voltage (pressure fixed to 2 mTorr) on the boundary interaction of RFD: (a1) and (a2) is time-averaged flux at the electrode surface, (b1) and (b2) is the mean energy of particles bombarding the electrode}
      \label{RFDFluxEnergyDiffPV}
    \end{figure}

    Even though RFD cannot be sustained, particle gain and loss are still balanced when they reach the periodic phase after being diagnosed over a longer time. In the positive discharge of argon, electrons come mainly from the ionization collision in the gap and SEE from the electrode surface and are lost at the boundary. Thus, for electrons, the boundary flux is mainly related to the ionization rate and SEE rate.
    
    From 0.1 mTorr to 3 mTorr, the discharge transforms from multipactor to RFD. In the RFD discharge region (0.5 - 3mTorr)The formation of an unstable sheath decreases the SEE rate, thus the electron flux decreases with the growth of gas pressure, as shown in figure \ref{RFDFluxEnergyDiffPV}(a1). This also shows that the source of electrons in RFD is dominated by SEE, which corresponds to figure \ref{RunawayParticleBalance}(c).
    For ions, the flux is mainly related to the ionization rate; thus, the ion flux of the RFD always increases with an increase of gas pressure. In the plasma of rf, the ionization rate does not simply increase with increasing gas pressure, but is also affected by the heating mode\cite{lieberman2005principles,liu2011collisionless,schmidt2015stochastic}. The effect of electromagnetic field on plasma makes the relationship between particle flux and gas pressure no longer simple.    
    
\begin{figure}[ht]
       \centering
         \includegraphics[width=0.9\textwidth]{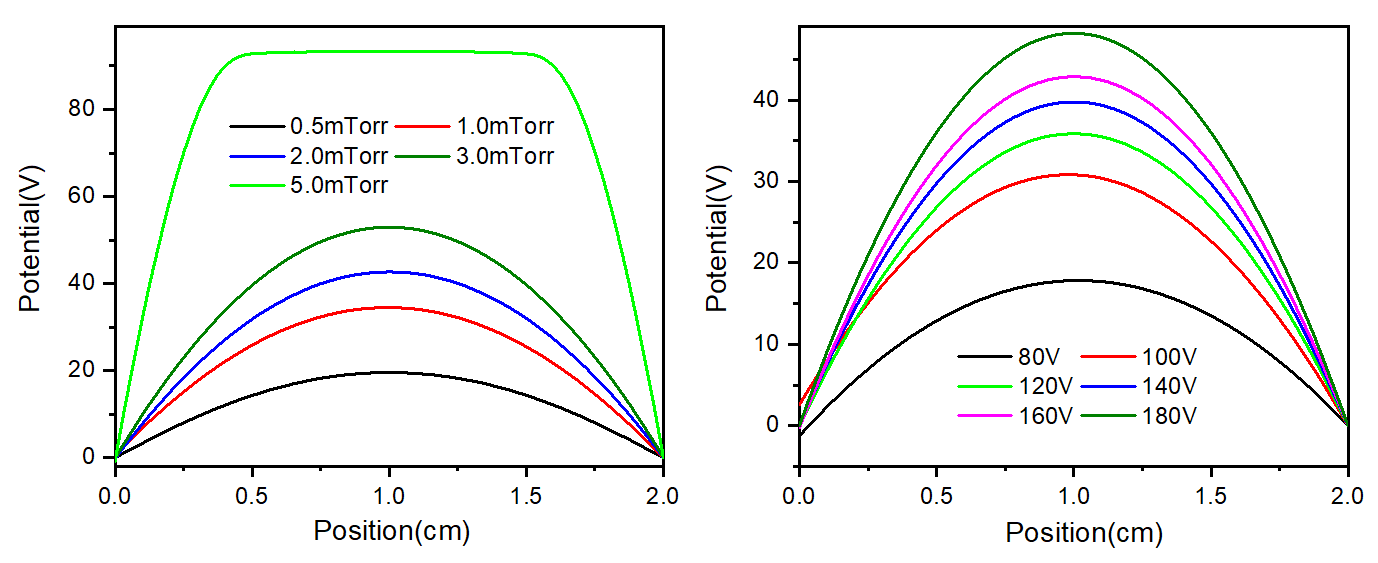}
         \caption{The distribution of time-averaged potential under different pressure (voltage fixed to 160 V) and voltage (pressure fixed to 2 mTorr) }
      \label{RFDPotentialDiffPV}
    \end{figure}

     The energy of the particles bombarding the electrode is mainly affected by the structure of the potential, which is shown in figure \ref{RFDPotentialDiffPV}. 
     When the gas pressure is lower than 3 mTorr, the electric field rf can fully penetrate the whole discharge gap, since the electron density is too low. Under the action of a 60 MHz oscillating penetrating electric field, the electrons in the entire gap maintain an average energy of tens of electron volts, which also dominate the energy of electrons bombarding the electrode, as shown in Figures \ref{RFDFluxEnergyDiffPV}(b1) and (b2).
     
     However, the accumulation of ions caused by ionization collisions can still form a positive potential barrier that will bind the electrons and accelerate the ions to bombard the electrodes. The center potential is still affected by the ionization rate, which is modulated by the gas pressure. Therefore, in the case of RFD and multipactor, under the force of increasing potential, the electron energy bombards the electrode and gradually decreases with increasing potential or gas pressure. The energy of the ions at the boundary is the opposite. Under the acceleration of the average electric field, the energy gradually increases with increasing gas pressure.

     The formation of glow discharge builds a strong sheath electric field that blocks electrons and speeds up ions from moving to electrodes. The existence of the self-generated field makes it difficult for the electric field to penetrate the bulk region, making it difficult for electrons to gain energy, so the electron energy of the glow discharge bombardment electrodes is much lower than RFD and multipactor.
     
     At the same time, the mean free paths of electrons and ions gradually shorten and become progressively non-negligible, which is the main reason for the decrease in the energy of the ions bombarding the electrode as the gas pressure increases. Under the force of increasing background gas pressure, the energy of the ions bombarding the electrode begins to decrease with an increase in gas pressure after a brief increase.

     The central potential is always positively related to the voltage amplitude, no matter whether for glow discharge, RFD, or multipactor, just as shown in figure \ref{RFDPotentialDiffPV}(b). For RFD, even though the increases in central potential will decrease the energy of electrons bombarding the electrode. Under acceleration of the applied rf voltage of more than 100 volts, the deceleration force on the electron from the self-generated potential of tens of volts of RFD can be almost ignored. 
     
     However, the 60 MHz oscillating rf field has no effect on ions because of its much higher mass. Tens of volts of the average self-generated potential can significantly accelerate the ions from the gap to the electrodes. Therefore, in RFD, a higher rf voltage directly results in a higher energy and higher flux for electrons and ions bombarding the electrodes, just as shown in Figures \ref{RFDFluxEnergyDiffPV} (b1) and (b2). Because of low gas pressure, there is almost no collision before the ion bombards the plate, so it has good anisotropy, which may be a reference value for the improvement of some special processes (such as high-aspect-ratio etching). At the same time, high boundary flux and electron energy can also alleviate the charging effect.

\section{Conclusion}

   In this work, we used a one-dimensional implicit PIC/MCC model to study the discharge of CCP under extremely low pressure driven by high-frequency rf power in pure argon. The relatively complete electron-induced SEE model that is suitable for electrodes covered with silicon dioxide and aluminum oxide films is considered. We approximate the external circuit by inserting a blocking capacitor between the rf power and the powered electrode. We found that there may be several types of unsustainable stable discharge modes at the left end of Paschen's curve. 
   
   We show and analyze several reasons for the formation of unsustainable discharges, including normal failure discharge caused by higher and lower rf voltages, runaway failure discharge caused by insufficient SEE, and bias failure discharge caused by the charge effect of the blocking capacitor. When the seed electrons are introduced again after the original electrons escape, the electron avalanche will occur again in runaway failure breakdown and the escape discharge caused by insufficient SEE will produce a periodic runaway failure discharge.
   
   We then explored and discussed the discharge characteristics of the runaway failure discharge. The electron flux at the electrode is extremely high and even more than in the glow discharge; The energy of electrons bombarding the electrode is much higher than in the case of the glow discharge. which may be useful for some special industrial processes. Note that all the results and discussions are drawn based on the 1D PIC/MCC code, which means the geometric effects that are important in many cases are neglected. We will verify them in 2D and 3D code in future works.

%\clearpage

\section*{Acknowledgments}
This work was supported by the National Natural Science Foundation of China (12275095, 11975174 and 12011530142), and the Fundamental Research Funds for Central Universities (WUT: 2020IB023), and the Hubei Provincial Natural Science Foundation of China (2023AFB488),
and Hubei University of Science and Technology Doctoral Startup Foundation (BK202401).

\section*{Data availability statement}
The data that support the findings of this study are available from the corresponding author upon reasonable request.

\section*{ORCIDs}
Hao Wu https://orcid.org/0000-0003-1074-6853\\
Ran An https://orcid.org/0000-0003-2692-401X\\
Wei Jiang https://orcid.org/0000-0002-9394-585X\\
Ya Zhang https://orcid.org/0000-0003-0473-467X\\

\nolinenumbers

%This is where your bibliography is generated. Make sure that your .bib file is actually called library.bib
\bibliography{library}

\providecommand{\noopsort}[1]{}\providecommand{\singleletter}[1]{#1}%
\begin{thebibliography}{10}

\bibitem{lieberman2005principles}
Michael~A Lieberman and Alan~J Lichtenberg.
\newblock {\em Principles of plasma discharges and materials processing}.
\newblock John Wiley \& Sons, 2005.

\bibitem{chen2023note}
Lei Chen, Hao Wu, Zili Chen, Yu~Wang, Lin Yi, Wei Jiang, and Ya~Zhang.
\newblock Note on particle balance in particle-in-cell/monte carlo model and
  its implications on the steady-state simulation.
\newblock {\em Plasma Sources Science and Technology}, 32(3):034001, 2023.

\bibitem{lisovskiy2008similarity}
V~Lisovskiy, J-P Booth, K~Landry, D~Douai, V~Cassagne, and V~Yegorenkov.
\newblock Similarity law for rf breakdown.
\newblock {\em Europhysics Letters}, 82(1):15001, 2008.

\bibitem{fu2020electrical}
Yangyang Fu, Peng Zhang, John~P Verboncoeur, and Xinxin Wang.
\newblock Electrical breakdown from macro to micro/nano scales: a tutorial and
  a review of the state of the art.
\newblock {\em Plasma Research Express}, 2(1):013001, 2020.

\bibitem{jin2022particle}
Biemeng Jin, Jian Chen, Alexander~V Khrabrov, Zhibin Wang, and Liang Xu.
\newblock Particle-in-cell simulations of the direct-current argon breakdown
  process in the 10--300 kv range.
\newblock {\em Plasma Sources Science and Technology}, 31(11):115015, 2022.

\bibitem{jiang2022gas}
Wei Jiang, Hao Wu, Zhijiang Wang, Lin Yi, and Ya~Zhang.
\newblock Gas breakdown in radio-frequency field within mhz range: a review of
  the state of the art.
\newblock {\em Plasma Science and Technology}, 2022.

\bibitem{banna2012pulsed}
Samer Banna, Ankur Agarwal, Gilles Cunge, Maxime Darnon, Erwine Pargon, and
  Olivier Joubert.
\newblock Pulsed high-density plasmas for advanced dry etching processes.
\newblock {\em Journal of Vacuum Science \& Technology A: Vacuum, Surfaces, and
  Films}, 30(4):040801, 2012.

\bibitem{agarwal2012extraction}
Ankur Agarwal, Shahid Rauf, and Ken Collins.
\newblock Extraction of negative ions from pulsed electronegative capacitively
  coupled plasmas.
\newblock {\em Journal of Applied Physics}, 112(3):033303, 2012.

\bibitem{adamovich20172017}
Igor Adamovich, SD~Baalrud, Anemie Bogaerts, PJ~Bruggeman, M~Cappelli, Vittorio
  Colombo, Uwe Czarnetzki, Ute Ebert, JG~Eden, Pietro Favia, et~al.
\newblock The 2017 plasma roadmap: Low temperature plasma science and
  technology.
\newblock {\em Journal of Physics D: Applied Physics}, 50(32):323001, 2017.

\bibitem{geng2015experimental}
Zhenxin Geng, Xin Lin, Jianyuan Xu, Xuebin Li, Xuchen Lu, and Zhuangzhang Yang.
\newblock Experimental study of cf4 insulation performance.
\newblock In {\em 2015 3rd International Conference on Electric Power
  Equipment--Switching Technology (ICEPE-ST)}, pages 394--397. IEEE, 2015.

\bibitem{zhang2016experiment}
Jia Zhang, Xin Lin, An~Su, Huili CHEN, and Luwei LI.
\newblock Experiment on breakdown characteristics of sf6/cf4 gas mixture under
  two typical electrode arrangements [j].
\newblock {\em High Voltage Apparatus}, 52(12):93--98, 2016.

\bibitem{seeger2017breakdown}
Martin Seeger, Patrick Stoller, and Angelos Garyfallos.
\newblock Breakdown fields in synthetic air, co 2, a co 2/o 2 mixture, and cf 4
  in the pressure range 0.5--10 mpa.
\newblock {\em IEEE Transactions on Dielectrics and Electrical Insulation},
  24(3):1582--1591, 2017.

\bibitem{fu_effect_2017}
Yangyang Fu, Shuo Yang, Xiaobing Zou, Haiyun Luo, and Xinxin Wang.
\newblock Effect of distribution of electric field on low-pressure gas
  breakdown.
\newblock {\em Physics of Plasmas}, 24(2):023508, 2017.
\newblock 00000.

\bibitem{fu_electrical_2020}
Yangyang Fu, Peng Zhang, John~P Verboncoeur, and Xinxin Wang.
\newblock Electrical breakdown from macro to micro/nano scales: a tutorial and
  a review of the state of the art.
\newblock {\em Plasma Research Express}, 2(1):013001, 2020.

\bibitem{hohn1997transition}
F~H{\"o}hn, W~Jacob, R~Beckmann, and R~Wilhelm.
\newblock The transition of a multipactor to a low-pressure gas discharge.
\newblock {\em Physics of Plasmas}, 4(4):940--944, 1997.

\bibitem{udiljak2003new}
Richard Udiljak, D~Anderson, P~Ingvarson, U~Jordan, U~Jostell, L~Lapierre,
  G~Li, M~Lisak, J~Puech, and J~Sombrin.
\newblock New method for detection of multipaction.
\newblock {\em IEEE transactions on plasma science}, 31(3):396--404, 2003.

\bibitem{wen2022higher}
De-Qi Wen, Peng Zhang, Janez Krek, Yangyang Fu, and John~P Verboncoeur.
\newblock Higher harmonics in multipactor induced plasma ionization breakdown
  near a dielectric surface.
\newblock {\em Physical Review Letters}, 129(4):045001, 2022.

\bibitem{zhang2019suppression}
Jianwei Zhang, Wei Luo, Hongguang Wang, Chunliang Liu, Yongdong Li, and Shu
  Lin.
\newblock Suppression of high-power microwave window breakdown by the
  sweeping-out-electron effect with an external dc bias electric field.
\newblock {\em Physics of Plasmas}, 26(12), 2019.

\bibitem{iqbal2023two}
Asif Iqbal, De-Qi Wen, John Verboncoeur, and Peng Zhang.
\newblock Two surface multipactor with non-sinusoidal rf fields.
\newblock {\em Journal of Applied Physics}, 134(15), 2023.

\bibitem{sato1997breakdown}
Masumi Sato and Masafumi Shoji.
\newblock Breakdown characteristics of rf argon capacitive discharge.
\newblock {\em Japanese journal of Applied Physics}, 36(9R):5729, 1997.

\bibitem{lisovskiy1998rf}
VA~Lisovskiy and VD~Yegorenkov.
\newblock Rf breakdown of low-pressure gas and a novel method for determination
  of electron-drift velocities in gases.
\newblock {\em Journal of Physics D: Applied Physics}, 31(23):3349, 1998.

\bibitem{vender1996simulations}
D~Vender, HB~Smith, and RW~Boswell.
\newblock Simulations of multipactor-assisted breakdown in radio frequency
  plasmas.
\newblock {\em Journal of applied physics}, 80(8):4292--4298, 1996.

\bibitem{wu2021electrical}
Hao Wu, Youyou Zhou, Jiamao Gao, Yanli Peng, Zhijiang Wang, and Wei Jiang.
\newblock Electrical breakdown in dual-frequency capacitively coupled plasma: A
  collective simulation.
\newblock {\em Plasma Sources Science and Technology}, 2021.

\bibitem{lisovskiy2005extinction}
Valeriy Lisovskiy, J-P Booth, Sofia Martins, Karine Landry, David Douai, and
  Valerick Cassagne.
\newblock Extinction of rf capacitive low-pressure discharges.
\newblock {\em Europhysics Letters}, 71(3):407, 2005.

\bibitem{horvath2017role}
B~Horv{\'a}th, M~Daksha, I~Korolov, A~Derzsi, and J~Schulze.
\newblock The role of electron induced secondary electron emission from sio2
  surfaces in capacitively coupled radio frequency plasmas operated at low
  pressures.
\newblock {\em PLASMA SOURCES SCIENCE \& TECHNOLOGY}, 26(12), 2017.

\bibitem{horvath2018effect}
Benedek Horv{\'a}th, Julian Schulze, Zolt{\'a}n Donk{\'o}, and Aranka Derzsi.
\newblock The effect of electron induced secondary electrons on the
  characteristics of low-pressure capacitively coupled radio frequency plasmas.
\newblock {\em Journal of Physics D: Applied Physics}, 51(35):355204, 2018.

\bibitem{kim2006transition}
HC~Kim and JP~Verboncoeur.
\newblock Transition of window breakdown from vacuum multipactor discharge to
  rf plasma.
\newblock {\em Physics of plasmas}, 13(12):123506, 2006.

\bibitem{na2019analysis}
Dong-Yeop Na and Fernando~L Teixeira.
\newblock Analysis of multipactor effects by a particle-in-cell algorithm
  integrated with secondary electron emission model on irregular grids.
\newblock {\em IEEE Transactions on Plasma Science}, 47(2):1269--1278, 2019.

\bibitem{hubble2017multipactor}
Aimee~A Hubble, Vernon~H Chaplin, Kathryn~A Clements, Rostislav Spektor,
  Preston~T Partridge, and Timothy~P Graves.
\newblock Multipactor breakdown threshold reduction due to magnetic confinement
  in parallel fields.
\newblock {\em IEEE Transactions on Plasma Science}, 45(7):1726--1730, 2017.

\bibitem{spektor2018space}
R~Spektor, MS~Feldman, AA~Hubble, and TP~Graves.
\newblock Space charge saturation in multipactor discharges with parallel
  magnetic field.
\newblock {\em Physics of Plasmas}, 25(12):122109, 2018.

\bibitem{feldman2018effects}
Matthew~S Feldman, Aimee~A Hubble, Rostislav Spektor, and Preston~T Partridge.
\newblock Effects of backscattered electrons on multipactor simulations with
  parallel magnetic fields.
\newblock In {\em 2018 IEEE MTT-S International Conference on Numerical
  Electromagnetic and Multiphysics Modeling and Optimization (NEMO)}, pages
  1--3. IEEE, 2018.

\bibitem{iqbal2023recent}
Asif Iqbal, De-Qi Wen, John Verboncoeur, and Peng Zhang.
\newblock Recent advances in multipactor physics and mitigation.
\newblock {\em High Voltage}, 2023.

\bibitem{verboncoeur2005particle}
John~P Verboncoeur.
\newblock Particle simulation of plasmas: review and advances.
\newblock {\em Plasma Physics and Controlled Fusion}, 47(5A):A231, 2005.

\bibitem{guo2019secondary}
Junjiang Guo, Dan Wang, Yantao Xu, Xiangping Zhu, Kaile Wen, Guanghui Miao,
  Weiwei Cao, JinHai Si, Min Lu, and Haitao Guo.
\newblock Secondary electron emission characteristics of al2o3 coatings
  prepared by atomic layer deposition.
\newblock {\em AIP advances}, 9(9):095303, 2019.

\bibitem{smith2003breakdown}
HB~Smith, Christine Charles, and RW~Boswell.
\newblock Breakdown behavior in radio-frequency argon discharges.
\newblock {\em Physics of Plasmas}, 10(3):875--881, 2003.

\bibitem{vahedi1993capacitive}
V~Vahedi, G~DiPeso, CK~Birdsall, MA~Lieberman, and TD~Rognlien.
\newblock Capacitive rf discharges modelled by particle-in-cell monte carlo
  simulation. i. analysis of numerical techniques.
\newblock {\em Plasma Sources Science and Technology}, 2(4):261, 1993.

\bibitem{kawamura2000physical}
E~Kawamura, Charles~K Birdsall, and Vahid Vahedi.
\newblock Physical and numerical methods of speeding up particle codes and
  paralleling as applied to rf discharges.
\newblock {\em Plasma Sources Science and Technology}, 9(3):413, 2000.

\bibitem{wang2010implicit}
Hong-yu Wang, Wei Jiang, and You-nian Wang.
\newblock Implicit and electrostatic particle-in-cell/monte carlo model in
  two-dimensional and axisymmetric geometry: I. analysis of numerical
  techniques.
\newblock {\em Plasma Sources Science and Technology}, 19(4):045023, 2010.

\bibitem{vahedi1995monte}
Vahid Vahedi and Maheswaran Surendra.
\newblock A monte carlo collision model for the particle-in-cell method:
  applications to argon and oxygen discharges.
\newblock {\em Computer Physics Communications}, 87(1-2):179--198, 1995.

\bibitem{phelps1999cold}
AV~Phelps and Z~Lj Petrovic.
\newblock Cold-cathode discharges and breakdown in argon: surface and gas phase
  production of secondary electrons.
\newblock {\em Plasma Sources Science and Technology}, 8(3):R21, 1999.

\bibitem{townsend1915electricity}
John~Sealy Townsend.
\newblock {\em Electricity in gases}.
\newblock Ripol Classic, 1915.

\bibitem{korolov2014experimental}
Ihor Korolov, Aranka Derzsi, and Zolt{\'a}n Donk{\'o}.
\newblock Experimental and kinetic simulation studies of radio-frequency and
  direct-current breakdown in synthetic air.
\newblock {\em Journal of Physics D: Applied Physics}, 47(47):475202, 2014.

\bibitem{puavc2018monte}
Marija Pua{\v{c}}, Dragana Mari{\'c}, Marija Radmilovi{\'c}-Radjenovi{\'c},
  Milovan {\v{S}}uvakov, and Zoran~Lj Petrovi{\'c}.
\newblock Monte carlo modeling of radio-frequency breakdown in argon.
\newblock {\em Plasma Sources Science and Technology}, 27(7):075013, 2018.

\bibitem{puavc2020monte}
Marija Pua{\v{c}}, Antonije Devi{\'c}, and Zoran~Lj Petrovi{\'c}.
\newblock Monte carlo simulation of rf breakdown in oxygen--the role of
  attachment.
\newblock {\em The European Physical Journal D}, 74:1--8, 2020.

\bibitem{radmilovic2005modeling}
M~Radmilovi{\'c}-Radjenovi{\'c} and JK~Lee.
\newblock Modeling of breakdown behavior in radio-frequency argon discharges
  with improved secondary emission model.
\newblock {\em Physics of plasmas}, 12(6):063501, 2005.

\bibitem{liu2011collisionless}
Yong-Xin Liu, Quan-Zhi Zhang, Wei Jiang, Lu-Jing Hou, Xiang-Zhan Jiang, Wen-Qi
  Lu, and You-Nian Wang.
\newblock Collisionless bounce resonance heating in dual-frequency capacitively
  coupled plasmas.
\newblock {\em Physical review letters}, 107(5):055002, 2011.

\bibitem{schmidt2015stochastic}
Christian Schmidt and Alexander Piel.
\newblock Stochastic heating of a single brownian particle by charge
  fluctuations in a radio-frequency produced plasma sheath.
\newblock {\em Physical Review E}, 92(4):043106, 2015.

\end{thebibliography}

%This defines the bibliographies style. Search online for a list of available styles.
\bibliographystyle{unsrt}

\end{document}